\documentclass[12pt,a4paper]{article}
\usepackage{amsmath,amssymb}
\usepackage{graphicx,epsfig}
\usepackage{color}

\newcommand{\gsim}{\, \mathop{}_{\textstyle \sim}^{\textstyle >} \,}
\newcommand{\lsim}{\, \mathop{}_{\textstyle \sim}^{\textstyle <} \,}

\def\vev#1{\left\langle#1\right\rangle}

\font\tenrsfs=rsfs10 at 12pt
\font\sevenrsfs=rsfs7
\font\fiversfs=rsfs5
\newfam\rsfsfam
\textfont\rsfsfam=\tenrsfs
\scriptfont\rsfsfam=\sevenrsfs
\scriptscriptfont\rsfsfam=\fiversfs
\def\mathscr#1{{\fam\rsfsfam\relax#1}}
\def\Lag{\mathscr{L}}

\begin{document}\thispagestyle{empty}

\begin{flushright}
CERN-PH-TH/2005-080\\
IFUP--TH/2005--12\\
UCB-PTH-05/15\\
LBNL-57568
\end{flushright}
\vskip 1.0cm

\begin{center}{\LARGE \bf Dark Energy and Right-Handed Neutrinos}

   \vskip 1.0cm

  {\large Riccardo Barbieri$^{a,b}$, Lawrence J.~Hall$^c$,
   Steven J.~Oliver$^c$, Alessandro Strumia$^d$}\\[1cm]
{\it $^a$ Scuola Normale Superiore and INFN, Piazza dei Cavalieri 7, I-56126 Pisa, Italy} \\[5mm]
{\em ${}^b$Theoretical Physics Division, CERN, CH-1211 Gen\`eve 23, Switzerland}\\[5mm]
{\it $^c$ Department of Physics, University of California, Berkeley, and\\
  Theoretical Physics Group, LBNL,
  Berkeley, CA 94720, USA}\\[5mm]
{\it $^d$ Dipartimento di Fisica dell'Universit{\`a} di Pisa and INFN, Italia}\\[3mm]

   \vskip 1.0cm

{\large\bf Abstract}
\begin{quote}\large
We explore the possibility
that a CP violating phase of the neutrino mass matrix is promoted to a
pseudo-Goldstone-boson field and is identified as the quintessence field for Dark Energy.
By requiring that the quintessence potential be calculable from a Lagrangian,
and that the  extreme flatness of the potential be stable under
radiative corrections, we are led to an essentially unique model.
Lepton number is violated only by Majorana masses of
light  right-handed neutrinos, comparable
to the Dirac masses that mix right- with left-handed neutrinos.
We outline the rich and constrained neutrino phenomenology that results from this proposal.
\end{quote}
\end{center}
\newpage


\section{Introduction}
\label{sec:intro}

A series of different observations and considerations \cite{data, review}
provides a strong case for a striking phenomenon: 
the expansion of the universe has recently begun to accelerate. 
Although a definitive experiment with sufficiently small systematic
uncertainties is lacking, 
if confirmed this remarkable fact calls for 
an adequate explanation and, even more important, motivates the search for 
other correlated observable phenomena. 

The accelerated expansion of the universe could be due to a tiny Cosmological 
Constant (CC), $\Lambda \approx (3 \cdot 10^{-3} \mbox{eV})^4$; tiny, but 
non-zero. In fact, the frustration generated by the unsuccessful attempts 
to solve the vacuum energy problem has led to the development of a mild 
anthropic interpretation of the apparently observed value of the CC 
\cite{Weinberg}. Here 
we take the view that the search for a more fundamental understanding of 
the cosmic acceleration remains highly motivated, even if still 
resting on the assumption of an exactly vanishing vacuum energy.

As an alternative to a CC, the accelerated expansion of the universe may be due to the 
evolution of some scalar field,  uniform or quasi-uniform in space, 
with the associated ``Dark Energy" (DE) mostly in its potential, 
usually called ``quintessence" \cite{Peebles:1987ek}. 
Signals related to such an interpretation,  although typically only vaguely 
determined, could be an equation of state of the 
associated fluid different from the one of a pure CC, or the effects of 
the couplings of the quintessence field to the usual matter or gauge fields. 

\medskip

In this work we describe a possible microscopic origin for the quintessence 
field and for its potential, guided by two general requirements and 
one phenomenological observation. One general requirement is that the 
quintessence potential should be calculable from a Lagrangian and its peculiar 
properties, in particular its extreme flatness, should be stable under 
radiative corrections. Since the mass scale governing this flatness is 
todays Hubble parameter, $H_0 \approx 10^{-33}$ eV, this requirement is 
severe indeed, although it is known to be satisfied by a Pseudo-Goldstone 
Boson (PGB) arising from the spontaneous breaking of a global symmetry near 
the Planck scale \cite{Weiss:1987xa}. 
Our second general requirement is that the physics of DE be 
directly connected to observable particle physics, so that the resulting theory 
can be tested in the laboratory. These two requirements appear to conflict --- if 
the quintessence field is coupled sufficiently strongly to the standard model 
to give laboratory signals, then radiative corrections involving this 
coupling will destroy the extreme flatness of the potential. A hint of a possible 
escape from this conundrum is provided by the phenomenological observation, 
already made by several people \cite{Frieman:1995pm,Nelson}, 
of the relative closeness of the energy scale associated with 
DE to the scale of neutrino masses. Thus we are led to explore the possibility 
that a CP violating phase of the neutrino mass matrix is promoted to a 
PGB field and is identified as the quintessence field for DE.


\section{The model}

To implement this idea we introduce right-handed neutrinos, $N_i$, at least two but most 
likely three, and as many complex scalars,  $\phi_{i j}$, as there are 
independent Lorentz-invariant neutrino bilinears $N_i N_j$. 
These scalars have Yukawa couplings to the $N_i$ 
$( \phi_{i j} = \phi_{j i}, \lambda_{i j} = \lambda_{j i})$
\begin{equation}
   \Lag_Y^N = \frac{1}{2} \sum_{i j} \lambda_{i j} \; \phi_{i j} N_iN_j,
\label{eq:LN}
\end{equation}
which are invariant under independent phase transformations 
of each $N_i$ field, say ${\rm U(1)}^3$ for concreteness. ${\rm U(1)}^3$ is 
a subgroup of the ${\rm U(1)}^6$ which transforms each of the $\phi_{i j}$-fields 
by an independent phase. The crucial assumption is that, in the limit 
of vanishing $\lambda_{i j}$, the full Lagrangian is invariant under this  
global ${\rm U(1)}^6$, which is spontaneously broken by the vacuum expectation 
values  $\vev{\phi_{i j}} \equiv f_{ij}$.

In the absence of any other coupling of the $N_i$ fields, this model has three 
massless Goldstone bosons and three PGBs, $G_{ij}$. The effective potential for the
combinations of $G_{ij}$ that are PGBs arises at one loop and is given by
\begin{equation}
V_1 \approx \frac{1}{32 \pi^2} {\rm Tr}\bigg[ M M^\dagger M M^\dagger~
\ln{\frac{\Lambda^2}{M M^\dagger}}\bigg]
\label{eq:Vlog}
\end{equation}
where $M_{i j} = \lambda_{i j} f_{i j}e^{iG_{ij}/f_{ij}}$ is the 
field-dependent, right-handed 
neutrino mass matrix and $\Lambda$ is a cut-off, to be specified later. 
Note the irrelevance in $V$ of any quadratic term in $M_{i j}$, however 
generated, since the only such term invariant under ${\rm U(1)}^3$ is also 
${\rm U(1)}^6$-invariant, and therefore $G_{ij}$ independent. 
A typical term in $V_1$ contributing to the potential of a PGB field, 
$G$, has the form 
\begin{equation}
V(G) = \mu^4 \cos{(G/f)}
\label{eq:VPGB}
\end{equation}
where $\mu^4 = O(M^4)$ arises as a product $M_{ij}M_{jk}^*M_{kl}M_{li}^*$,
and $f$ is an appropriate function of the symmetry breaking 
parameters $f_{ij}$. It is well-known that, with $\mu \approx 3 \times 10^{-3}$ eV and 
$f$ of order $M_{\rm Pl}$, $G$ is a consistent candidate for the 
quintessence field \cite{Weiss:1987xa,Frieman:1995pm,PGBDE}.
However, the signals we wish to stress are not those that come from the form of 
the potential (\ref{eq:VPGB}), but rather are due to the required form for the 
underlying neutrino sector.

\smallskip

Two natural and important questions arise at this point. Could we interpret 
the $N_i$  as the left-handed neutrinos entering the usual left-handed lepton 
doublets $L_i$? What other couplings can complete consistently the neutrino 
sector? To answer the first question we should first transform eq. (\ref{eq:LN}) into a 
gauge invariant interaction involving the $L_i$ and the Higgs doublet $h$
\begin{equation}
   \Lag^L_Y =  \frac{1}{2} \sum_{i j} \lambda^L_{i j} \; \phi_{i j}  
\frac{ h^2}{M_L^2} L_iL_j,
\label{eq:LL}   
\end{equation}
where gauge indices are left understood and $M_L$ is an energy scale introduced 
to give   $\Lag^L_Y $ the correct dimensions. Indeed, if we now replace 
the Higgs field with its vacuum expectation value, we would be led to the same 
contribution as in (\ref{eq:Vlog}) to the PGB potential, except that  $M_{i j}$ 
would now be the left-handed neutrino mass matrix. Apparently this is the minimal 
theory, with the DE field directly related to the phases of
the $3 \times 3$ neutrino  
mixing matrix. It is straightforward to see, however, 
that radiative corrections above the weak scale with internal Higgs fields 
destroy radiative stability. This is an important conclusion. Without considerable 
complications of the Higgs sector \cite{Hill:1988bu}, the introduction of light 
right-handed neutrinos appears as a necessity. A similar conclusion applies to 
the case with $\phi_{ij}$ fields coupled to the $L_i N_j$ bilinear.

We can now answer the second question stated above.
A consistent completion of the 
Yukawa Lagrangian in the lepton sector has the form 
\begin{equation}
   \Lag_Y =  \sum_{i j} \lambda^E_{i j} \; h L_i e^c_j
   + \sum_{i j} \lambda^D_{i j} \; h L_iN_j
+  \frac{1}{2} \sum_{i j} \lambda_{i j} \; \phi_{i j} N_iN_j
\label{eq:L}   
\end{equation}
involving the right-handed charged leptons, $e^c_i$. One 
recognizes the usual Dirac neutrino mass matrix, proportional to $\lambda^D$, 
and one notices the absence of any (gauge 
invariant) Majorana mass term for the left-handed neutrinos of the form 
 $ \lambda^M_{i j}  \,h^2 L_iL_j/M_L$. Such a term, in fact, would 
explicitly break lepton number and, in conjunction with the Dirac mass matrix, 
would allow terms in the PGB potential linear in $M_{i j}$, thus also destroying 
radiative stability. 
In fact, radiative stability requires that all non-renormalizable
operators conserve both ${\rm U(1)}^6$ on the $\phi_{ij}$ and an overall
lepton number, ${\rm U(1)}_L$.
In the presence of non-zero $\lambda_{i j}$, as well as generic $\lambda^{E,D}$
couplings, the ${\rm U(1)}^6$ symmetry is explicitly broken to ${\rm U(1)}_L$; it is 
convenient to label the theory by ``${\rm U(1)}^6 \rightarrow {\rm U(1)}_L$''.

\medskip

From the above arguments, the theory of (\ref{eq:L}) is essentially unique. The only fermion 
bilinear that $\phi_{ij}$ can couple to is $N_i N_j$, and the $h^2 L_i L_j$ operator 
must be absent. We conclude that there must be two or more light right-handed neutrinos, 
with typical entries in their Majorana mass matrix of order $3 \times 10^{-3}$ eV ---
broadly comparable to the entries in the Dirac matrix.
It is remarkable that, in promoting a CP violating phase of the neutino sector to a field,
the mass of this field can be protected to the level of $H_0 \approx 10^{-33}$ eV. 
The key is to ensure that the leading radiative correction to $m_G^2$ is of order 
$m_\nu^4/M_{\rm Pl}^2$. 

\medskip

It is the Dirac mass term in (\ref{eq:L}) that allows us to call $N_i$ the right-handed 
neutrinos. One may wonder whether this term introduces 
significant new corrections to the PGB potential. In fact it does at two-loop 
order, giving a term in the PGB potential 
\begin{equation}
V_2 \approx \left( \frac{1}{16 \pi^2} \right)^2 {\rm Tr}( M M^\dagger 
\lambda^D \lambda^{D \dagger})\,
\Lambda^2.
\label{eq:V2}
\end{equation}
This leads us to consider a supersymmetric 
extension of the model with spartner masses  at the Fermi scale, in which case a 
typical sneutrino mass cuts off both the quadratic divergence of this two-loop potential 
and the logarithmic divergence of the one-loop term in (\ref{eq:Vlog}). Up to loop factors, 
the resulting two-loop contribution also 
becomes of the relevant order of magnitude for a quintessence potential\footnote{In contrast
to the case of Mass Varying Neutrinos \cite{Nelson}, the
effective contribution to the quintessence potential from the cosmological neutrino 
density is numerically irrelevant.
We can also ignore the small variation of the neutrino masses  induced by 
their dependence on the dynamical PGB fields.}.
Note that with the complete $\Lag_Y$ all the would-be-Goldstones 
associated with the breaking of ${\rm U(1)}^6$ acquire a mass from the PGB potential, 
except for the linear combination related to the overall lepton number. 

We must be clear that we have not explained why the neutrinos are light --- 
quite the reverse, since we have introduced extremely small parameters in the 
$\lambda$ and $\lambda^D$ matrices. The PGB masses are small because they are 
proportional to powers of these small explicit symmetry breaking parameters. 
Neglecting flavor labels and emphasizing only the very small parameters, the interactions of 
(\ref{eq:L}) can be rewritten as $hLe^c + \varepsilon_D \,hLN + \varepsilon_M^2 \,\phi NN$, 
where $\varepsilon_{D,M}$ are now the small parameters.
These parameters may not be promoted to fields, with the lightness of the neutrinos 
explained in terms of a small vacuum value, since these fields would lead to
disastrous radiative corrections to the PGB potential.  Nevertheless, it is 
interesting that both $\varepsilon_D$ and $\varepsilon_M$ should take on values of order 
$10^{-13}$ to $10^{-15}$ for acceptable neutrino masses and DE ---
an approximate symmetry acts on the $N$ fields. In higher dimensional theories,
a small $\varepsilon$ could result if $N$ propagate in a bulk and have a small, 
exponentially suppressed wavefunction at the location of the $\phi,L$ and 
$h$ fields. Even in this case, any parameter that sets the geometrical configuration 
must not correspond to a light field in the low energy effective four dimensional theory.

\medskip

Two variations in the theory are possible: by restricting the form of 
the couplings or the number of $\phi$ fields in~(\ref{eq:L}), alternative 
symmetry breaking patterns emerge, yielding PGB potentials different from $V_1 + V_2$
of the generic case. If the entire theory possesses an exact
${\rm U(1)}^3_i$ symmetry, with one ${\rm U(1)}$ for each lepton generation, $\lambda^{E,D}$
become diagonal and we obtain the  ``${\rm U(1)}^6 \rightarrow {\rm U(1)}_i^3$'' variation. 
Since  $ {\rm Tr}\, (M M^\dagger \lambda^D  \lambda^{D \dagger})$ is now independent of the 
three PGB fields, the potential for the PGBs is given purely by $V_1$ of (\ref{eq:Vlog}).
As this potential has only a logarithmic divergence, the quintessence potential is 
stable to radiative corrections whether or not superpartners are at the weak scale.
In this variation, lepton flavor mixing arises entirely from spontaneous breaking.
Finally,  if the initial symmetry is restricted to ${\rm U(1)}_i^3$, so that the theory
possesses only three $\phi$ fields, $\phi_{ii}$, we obtain the 
``${\rm U(1)}^3_i \rightarrow {\rm U(1)}_L$'' variation. The potential for the 2 PGBs occurs at
3 loops
\begin{equation}
V_3 \approx \left( \frac{1}{16 \pi^2} \right)^3 
{\rm Tr}\,(M \lambda^{D \dagger} \lambda^D M^\dagger \lambda^D \lambda^{D \dagger}) \Lambda^2.
\label{eq:V3}
\end{equation}
This again gives a successful theory for quintessence with entries of $M$ of 
order $10^{-3}$ eV, but, in contrast to the general case, 
supersymmetry should be absent, giving a cutoff $\Lambda \approx M_{\rm Pl}$.
As the cutoff is reduced, so the entries of $M$ can be made larger --- with the cutoff
in this variation at the weak scale the right-handed neutrino masses 
may be raised to an MeV.

 
 \section{Constraints and Neutrino Spectra}
 
Since $f_{ij} \sim M_{\rm Pl}$ the PGB interactions are extremely weak, so that
the main consequences of our theory are in the neutrino mass sector.
With 3 right-handed neutrinos, the full neutrino mass matrix is $6 \times 6$ and 
is made of a Dirac mass matrix, $m_{i j} = \lambda^D_{i j} \vev{h}$, and of a Majorana 
mass matrix for the right-handed neutrinos, $M_{i j} = \lambda_{i j} f_{i j}$, both 
$3 \times 3$ and roughly of the same size (up to differences among the various 
matrix elements, which can of course be significant). An $M$ very much smaller 
or much bigger than $m$ would in fact give the wrong size for the PGB potential.
Such a neutrino mass matrix is mainly constrained by oscillation experiments. 
However, it is also constrained by cosmology: extra sterile neutrinos
coming into equilibrium, partially or totally, affects Big-Bang
Nucleosynthesis (BBN), the Cosmic Microwave Background (CMB) 
and Large Scale Structure (LSS) formation. 

\smallskip

After symmetry breaking, the neutrino masses and mixings can be described
in full generality by
\begin{equation}
\Lag = \frac{g}{\sqrt{2}} \bar{\nu} V \gamma_\mu  P_L e \, W_\mu + e^T m_E e^c
+ \nu^T m^d  N + \frac{1}{2} N^T U^T M^d U N + {\rm h.c.}
\label{eq:Lagr}
\end{equation}
where the flavour indices are left understood, $m_E, m^d$ and $M^d$ are real and diagonal matrices,
$V$ is a unitary matrix with a single physical phase and $U$ is a unitary matrix 
with five physical phases\footnote{The proof is as follows. The first three terms in the right-hand-side 
of~(\ref{eq:Lagr}) are the 
usual terms in the case of pure Dirac neutrinos, which can always be reduced to this form. The last term is an arbitrary symmetric matrix with the overall phase which is unphysical because it can be reabsorbed by an overall lepton number transformation.}.
If $m$ and $M$ are diagonalized by $m = V_D m^d U_D$ and $M = U_M^T 
M^d U_M$ respectively, then
$U = U_M U_D^\dagger$.
To analyze the constraints in full generality in the entire space 
of parameters is complicated and goes beyond the scope of this work. 
In the following we try to describe the main features of the allowed parameter 
space.

\smallskip

 \begin{figure}[t]
$$\includegraphics[width=17cm]{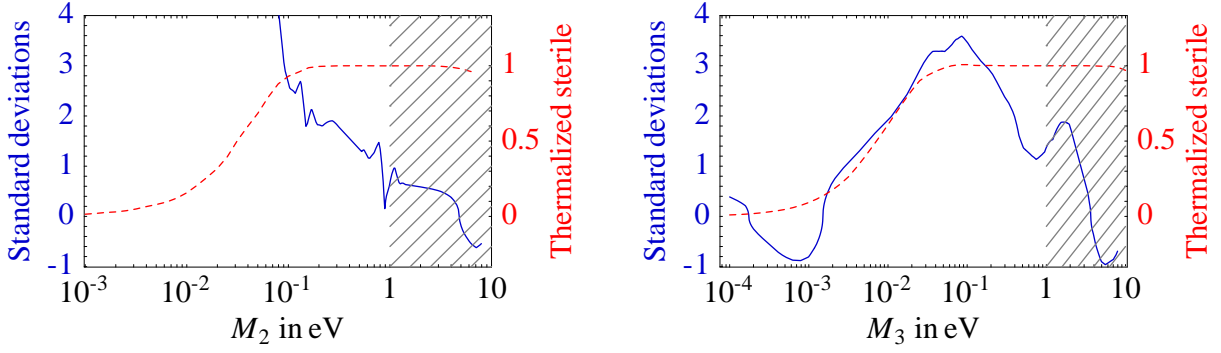}$$
\caption[X]{\label{fig:SunAtm}\em ${\rm Sign}(\Delta\chi^2)\cdot  |\Delta\chi^2|^{1/2}$
of the global oscillation fit (continuous blue line/left vertical axis)
and thermalized sterile fraction (red dotted line/right vertical axis).
Large Scale Structure data exclude thermalized heavy sterile neutrinos (shaded regions).
}
\end{figure}

We first consider the special case $U = 1$, in which $m$ and $M$ may
be simultaneously diagonalized. This
is a useful starting point to understand 
the more general situation or, at least,  to show that there are allowed regions in 
parameter space that fulfil all requirements. 
This situation is fully realistic and we study it below.

The constraints on the mass parameters are easy to determine because
diagonalization of the $6 \times 6$ neutrino mass matrix decouples
into 3 separate $2 \times 2$ diagonalizations, one for each ($\nu_i,
N_i$) pair. We
disregard possible degeneracies and order the eigenvalues of the 
Dirac and Majorana mass matrices, $m_i$ and $M_i$ respectively, 
in such a way that $m_3, M_3$ govern the atmospheric oscillation length and $m_2, M_2$
the solar oscillation length.  The constraints 
from oscillation experiments on $M_2$  and $M_3$  are shown in Fig.~\ref{fig:SunAtm}.  
Also shown are the respective fractions of thermalized 
sterile neutrinos at BBN and CMB eras, 
$\Delta N_\nu$, and the regions excluded by LSS data. An analogous 
figure cannot be drawn for $M_1$ since we do not know the mass of the lightest 
active, or quasi-active neutrino. However, for $m_1$ small enough, 
say $m_1 \lesssim 10^{-6}$ eV, $M_1$ is almost unconstrained.

\begin{figure}
\begin{center}
  \includegraphics[width=16cm]{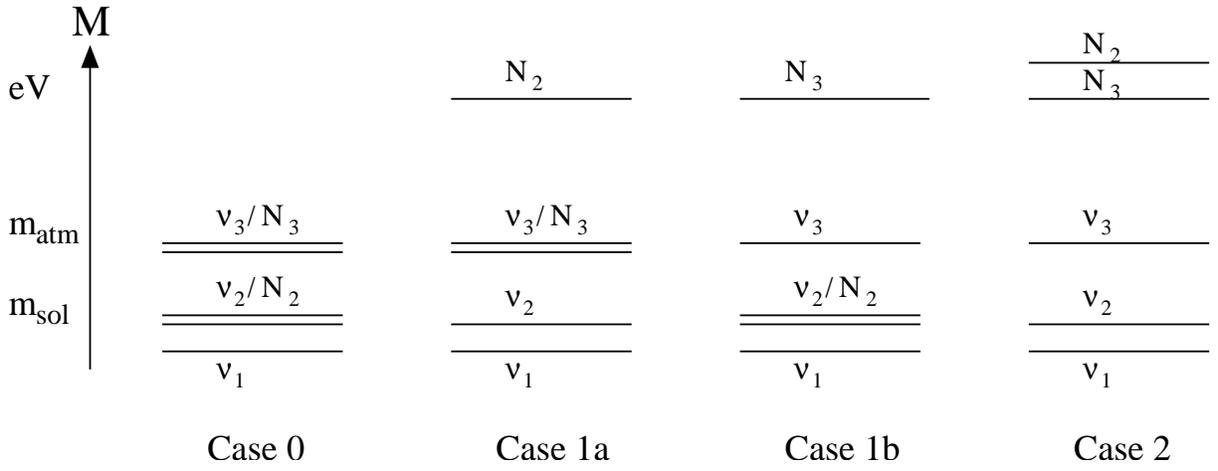}
\caption{\em Schematic illustration of four possible neutrino mass spectra consistent with
  oscillation and cosmological constraints. The mass of $N_1$ is largely undetermined.}
\label{fig:spectrum}
\end{center}
\end{figure}

While LSS forbids $M_{2,3} \gsim $eV, each of these masses could lie
in the ``0.3 eV window'', given by $0.1 \mbox{eV} \lsim M_{2,3} \lsim
\mbox{eV}$. Since the observed masses for atmospheric and solar
oscillations are less than 0.3 eV, these values for $M_{2,3}$ lead to
a mini-seesaw. Alternatively, although atmospheric oscillations exclude 
$M_3 \simeq m_3$, 
any value less than about $10^{-2}\,$eV is allowed, yielding a
pseudo-Dirac pair for $(\nu_3, N_3)$. On the other hand, solar
oscillations forbid $M_2$ beneath the 0.3 eV window all the way down to
$\sim 10^{-9}\,$eV. Allowed values below $10^{-9}$ eV lead to a
pseudo-Dirac $(\nu_2, M_2)$ pair.  Thus, each of  $(\nu_2, M_2)$ and
$(\nu_3, M_3)$ either undergo a mini-seesaw or form a pseudo-Dirac
pair. Furthermore, from  Fig.~\ref{fig:SunAtm} we see that the
cosmological thermalization of the sterile state is complete for the
mini-seesaw case and absent for the pseudo-Dirac case (except as $M_3$
approaches $10^{-2}$ eV, when partial thermalization occurs).
Hence we can identify three possibilities\footnote{Here $0.3$ eV 
means anywhere in the ``0.3 eV window'', $M_3 \approx
10^{-3}$ eV means any value of $M_3$ less than about $10^{-2}$ eV,
and  $M_2 \approx 10^{-9}$ eV means any value of $M_2$ less than about 
$10^{-9}$ eV.}  
\begin{enumerate}
\item[(0)] 
$\Delta N_\nu\approx0$: \;\;\;
$(M_2, M_3) \approx (10^{-9}, 10^{-3})$ eV.
\item[(1)] 
$\Delta N_\nu\approx1$: \;\;\;
(1a)\;\; $(M_2, M_3) \approx (0.3, 10^{-3})$ eV, or 
(1b)\;\; $(M_2, M_3) \approx (10^{-9}, 0.3)$ eV.
\item[(2)]  
$\Delta N_\nu\approx2$: \;\;\;
$(M_2, M_3) \approx (0.3, 0.3)$ eV.
\end{enumerate}
These four mass ranges correspond to the four
possible ways of assigning mini-seesaw and pseudo-Dirac spectra to
each of  $(\nu_2, M_2)$ and $(\nu_3, M_3)$, as shown in Fig.~\ref{fig:spectrum}. 
One extra neutrino at BBN looks compatible with standard cosmology, with systematic
effects taken into account, whereas two extra neutrinos appear definitely
problematic, unless one invokes an \emph{ad hoc} non standard cosmology,
such as large lepton
asymmetries or a MeV-scale reheating temperature.


Although we have set $U=1$, the unitary matrix $U_M = U_D$ is
completely undetermined by neutrino mass phenomenology --- the Euler
angles, $\theta$, of $U_M$ can be chosen to obtain the observed DE,
$\rho_{\rm DE}$. If $M_1$ is sufficiently small, the relevant entries
of  $M_{i j} = \lambda_{i j} f_{i j}$ are given by Euler angles
multiplied by $M_2$ or $M_3$. In cases (0), (1) and (2), $\rho_{\rm DE}$ typically
requires $\theta \approx 1, 10^{-2}$ and $10^{-4}$
respectively. Perhaps the case (0) is most natural, and, depending on
the precise value for $M_3$, could lead to an observable deviation of
$\Delta N_\nu$ from 0.  

Theories with $U \neq 1$ can be constructed that are both more natural
and more predictive than the $U=1$ case. In particular, suppose that, in
the original basis for $N$, we have the texture
\begin{equation}
M_{i j} = \lambda_{i j} f_{i j} = 
\left( \begin{array}{ccc} M_1 &\varepsilon_{12}  & \varepsilon_{13} \\
                        \varepsilon_{12} & M_2        & \varepsilon_{23} \\
                        \varepsilon_{13} & \varepsilon_{23} & M_3 \end{array} \right),
\;\;\;\;\;\;\;\;
m = \left( \begin{array}{ccc} m_1 &   0 &   0 \\
                                                      0 & m_2 & 0 \\
                                                      0 & 0   & m_3
                                                      \end{array}
                                                      \right).
\end{equation}
where the $\varepsilon_{ij}$ are either zero or sufficiently small that
the neutrino phenomenology is not significantly altered from the $U=1$ case.
The standard neutrino mixing angles arise from transformations on the
left-handed leptons, charged or neutral.
In each of the three cases described above, it is possible 
to introduce a single off-diagonal entry, $\varepsilon_{ij}$,
such that the appropriate PGB potential is generated,
$\rho_{\rm DE} \approx M_{ii}M_{ij}^*M_{jj}M_{ji}^*$ for some $(i,j)$. 

Realistic examples, corresponding to
cases (0) and (1) above, are 
  
\begin{enumerate}

\item[(0)]  
$(m_1, m_2, m_3) \approx (\lesssim 10^{-6} \mbox{eV}, m_{\rm sun}, m_{\rm atm})$, \;\;  \\
$(M_1, M_2, M_3) \approx (\gtrsim 10^{-3}, 10^{-9}, 10^{-3})$ eV, and 
$\varepsilon_{13} \approx 10^{-2}$ eV;

\item[(1)]  
$(m_1, m_2, m_3) \approx (0, 5 \times 10^{-3} \mbox{eV}, m_{\rm atm})$, \;\;  \\
$(M_1, M_2, M_3) \approx (0, 0.3, 10^{-8} \mbox{---}10^{-3})$ eV, and 
$\varepsilon_{23} \approx 10^{-4} \mbox{eV}\cdot (10^{-3} \mbox{eV}/M_3)^{1/2}$;

\end{enumerate}
with other  $\varepsilon_{ij}$ taken irrelevantly small.

\smallskip

\baselineskip 0.54cm
 
\section{Signals and Conclusions}

In promoting a CP violating phase of the neutrino mass matrix to
the DE quintessence field, we are led, essentially uniquely, to a
theory with light right-handed neutrinos, with possible spectra 
shown in Figure 2. 
This proposal can be tested in neutrino physics.
\begin{itemize}
\item The three cases with differing $\Delta N_\nu$ will easily be distinguished 
by precision CMB measurements at PLANCK, and perhaps at WMAP.
\item The range $M_2\sim 0.3$ eV can be completely tested
by cosmology (searching for sterile neutrino masses) and by
reactor experiments (searching for $\bar\nu_e$ disappearance
at base-line $\sim 10\,{\rm m}$).

\item Similarly, $M_3\sim 0.3$ eV can be 
tested by cosmology, and possibly by atmospheric neutrino experiments 
(HyperK, MONOLITH, IceCube) and beam experiments (MiniBoone, MINOS).

\item Long-baseline experiments will probe  $M_3\sim 10^{-3}\, {\rm eV}$.
 If $M_3$ approaches $10^{-2}\, {\rm eV}$, it can be determined by a CMB or BBN
measurement of $\Delta N_\nu$, and signals may appear in atmospheric oscillations.

\item Very small $M_{1,2,3}$ can give MSW resonances
in the sun and in supernov\ae,  as well as vacuum oscillations of
neutrinos  that travel cosmological distances.

\item MiniBoone  is currently testing the LSND anomaly. 
Constraints from other oscillation data disfavor its
interpretation in terms of sterile neutrinos, 
but do not fully exclude it.

\item The detection of a $0\nu2\beta$ signal would exclude this model, 
since the left-handed neutrinos do not have a direct Majorana mass
and the right-handed neutrinos are  light, so that
effects in $0\nu2\beta$ are suppressed by powers of $M/Q$
(where $Q\sim {\rm MeV}$ is the energy released in $0\nu2\beta$).

\end{itemize}
The Mass Varying Neutrino scheme for DE \cite{Nelson} 
also involves a light scalar coupled to neutrinos. The neutrino energy density 
plays a crucial role in the dynamics of DE because the scalar is both 
light and has a significant coupling to the neutrinos.  Thus the scheme predicts 
a characteristic shift in the position of the CMB peaks, corresponding to at 
least one species of neutrino scattering during the eV era \cite{CMBsignal}.
On the other hand, in our scheme the 
coupling of neutrinos to light PGBs is proportional to $M/f$, 
and is so small that all neutrinos free-stream during the CMB era.

We conclude  by noting other areas that are worth 
investigating. 
Since there are several PGBs, some might have masses
larger than todays Hubble parameter, so that they oscillate during the
recent evolution of the universe with 
characteristic signals related to the associated Jeans length~\cite{Zeldovich, Hu:2000ke}.
Could such a PGB
give all of the dark matter? The mass should be larger than
about $10^{-22}\,{\rm  eV}$, otherwise the uncertainty
principle prevents the formation of structures at sufficiently small scales~\cite{Hu:2000ke}.
Parameters exist that allow a unified
picture of both dark matter and DE.
In theories with sufficiently sparse textures, it may be possible to
compute the magnitude of the DE and dark matter energy densities from
measurements of neutrino masses and mixings.
Finally, the (super-)potential  that gives the $ \phi_{i j}$ 
a vev at a large scale could play a role in inflation:
a candidate for a such superpotential is
$W = \sum_{i j} S_{i j} (\sigma_{i j} \phi_{i j} \bar{\phi}_{i j} - \mu^2_{i j})$ where $S_{i j}, \phi_{i j}$ and $\bar{\phi}_{i j}$ are chiral supermultiplets.

\section*{Acknowledgments}
We thank D. Larson and Y. Nomura for many conversations.
This work is  supported  in part by MIUR and by the EU under RTN contract
MRTN-CT-2004-503369, by DOE under contracts
  DE-FG02-90ER40542 and DE-AC03-76SF00098 and by NSF grant
  PHY-0098840.


\begin{thebibliography}{99}
\bibitem{data}
  S.~Perlmutter {\it et al.}  [Supernova Cosmology Project Collaboration],
  Astrophys.\ J.\  {517}, 565 (1999) [astro-ph/9812133];
  A.~G.~Riess {\it et al.}  [Supernova Search Team Collaboration],
  Astron.\ J.\  {116}, 1009 (1998) [astro-ph/9805201];
  A.~G.~Riess {\it et al.}  [Supernova Search Team Collaboration],
  Astrophys.\ J.\  {607}, 665 (2004)  [astro-ph/0402512].
  
\bibitem{review}
  For a review see P.~J.~E.~Peebles and B.~Ratra,
  Rev.\ Mod.\ Phys.\  {75}, 559 (2003)
  [astro-ph/0207347].
  
\bibitem{Weinberg}
 S.~Weinberg,
  Phys.\ Rev.\ Lett.\  {59}, 2607 (1987);
 H.~Martel, P.~R.~Shapiro and S.~Weinberg,
  Astrophys.\ J.\  {492}, 29 (1998)
  [astro-ph/9701099].

\bibitem{Peebles:1987ek}
  P.~J.~E.~Peebles and B.~Ratra,
  Astrophys.\ J.\  {325}, L17 (1988);
  B.~Ratra and P.~J.~E.~Peebles,
  Phys.\ Rev.\ D {37}, 3406 (1988);
  C.~Wetterich,
  Nucl.\ Phys.\ B {302}, 668 (1988).

\bibitem{Weiss:1987xa}
  N.~Weiss,
  Phys.\ Lett.\ B {197}, 42 (1987).

\bibitem{Frieman:1995pm}
  J.~A.~Frieman, C.~T.~Hill, A.~Stebbins and I.~Waga,
  Phys.\ Rev.\ Lett.\  {75}, 2077 (1995).

\bibitem{Nelson}
R.~Fardon, A.E.~Nelson and N.~Weiner,
  JCAP {0410}, 005 (2004)
  [astro-ph/0309800].
  
\bibitem{PGBDE}
  J.E.~Kim,
  JHEP {9905}, 022 (1999) [hep-ph/9811509];
  JHEP {0006}, 016 (2000) [hep-ph/9907528];
  K.~Choi,
  Phys.\ Rev.\ D {62}, 043509 (2000) [hep-ph/9902292];
  Y.~Nomura, T.~Watari and T.~Yanagida,
  Phys.\ Lett.\ B {484}, 103 (2000); [hep-ph/0004182].
  J.~E.~Kim and H.P.~Nilles,
  Phys.\ Lett.\ B {553}, 1 (2003) [hep-ph/0210402];
L.J. Hall, Y. Nomura, S. J. Oliver,
[astro-ph/0503706].

\bibitem{Hill:1988bu}
  C.T.~Hill and G.~G.~Ross,
  Nucl.\ Phys.\ B {311}, 253 (1988).

 \bibitem{CMBsignal}
Z. Chacko, L.J. Hall, T. Okui, and S. Oliver
Phys.\ Rev.\ D {70}, (2004) 085008 [hep-ph/0312267].

\bibitem{Zeldovich}
M.~Khlopov, B.~Malomed and Ya.~B.~Zel'dovich,
Mon.\ Not.\ R.\ Astron.\ Soc.  {215}, 575 (1985).

\bibitem{Hu:2000ke}
  W.~Hu, R.~Barkana and A.~Gruzinov,
  Phys.\ Rev.\ Lett.\  {85}, 1158 (2000)
  [astro-ph/0003365].





\end{thebibliography}
\end{document}